\documentclass{aa}
\newcommand{\varcsec}{^{\prime\prime}}

\newcommand\ionn[2]{#1$\,${\textsc{#2}}}

\usepackage{amssymb}
\usepackage{textcomp}
\usepackage{graphicx}
\usepackage{epstopdf}
\usepackage{natbib}
\usepackage{float}
\usepackage{longtable}
\usepackage[bookmarks = true,colorlinks = true,linkcolor = blue,urlcolor = blue,citecolor = blue,bookmarks = true, linktocpage = true,hyperindex = true,breaklinks]{hyperref}
\usepackage{microtype}
\usepackage{color} 
\usepackage{xfrac} 
\usepackage{hyphenat}
\usepackage{inputenc}
\usepackage{fontenc}

\defcitealias{Joshi_2017}{Paper I}

\bibpunct{(}{)}{;}{a}{}{,}

\usepackage{txfonts}
\begin{document}

   \title{Three-dimensional magnetic structure of a sunspot: Comparison of the photosphere and upper chromosphere}
   
   \subtitle{}

  \author{Jayant Joshi\inst{\ref{inst1},\ref{inst2}}
           \and Andreas Lagg \inst{\ref{inst1}}
           \and Johann Hirzberger\inst{\ref{inst1}}
           \and Sami K. Solanki\inst{\ref{inst1},\ref{inst3}}}           
   \institute{Max-Planck-Institut f\"{u}r Sonnensystemforschung, Justus-von-Liebig-Weg 3,
              37077, G\"{o}ttingen, Germany\label{inst1} 
              \and Institute for Solar Physics, Department of Astronomy, Stockholm University, AlbaNova University Centre,
              SE-106 91 Stockholm, Sweden\label{inst2}
              \and School of Space Research, Kyung Hee University, Yongin, Gyeonggi Do, 446-701, 
              Republic of Korea\label{inst3}\\           
              \email{jayant.joshi@astro.su.se} 
              }
   \date{Received; accepted}

   \titlerunning{Three-dimensional magnetic structure of a sunspot}
   \authorrunning{Jayant Joshi et. al.}


 \abstract
  {}  
   {We investigate the magnetic field of a sunspot in the upper chromosphere and compare it to 
             the photospheric properties of the field.
    }
{We observed the main leading sunspot of the active region NOAA 11124 during two days with the Tenerife Infrared 
Polarimeter-2 (TIP-2) mounted at the German Vacuum Tower Telescope (VTT). Through inversion of Stokes spectra of the 
\ionn{He}{i}\, triplet at 10830\,\AA, we obtained the magnetic field vector of the upper chromosphere. For comparison with the photosphere,
we applied height-dependent inversions of the \ionn{Si}{i}\,10827.1\,\AA\, and \ionn{Ca}{i}\,10833.4\,\AA\ lines.}
{We found that the umbral magnetic field strength in the upper 
chromosphere is lower by a factor of 1.30-1.65 compared to the photosphere.
The magnetic field strength of the umbra decreases from the photosphere toward the upper chromosphere
by an average rate of 0.5-0.9\,G\,km$^{-1}$.
The difference in the magnetic field strength between both atmospheric 
layers steadily decreases from the sunspot center to the outer boundary 
of the sunspot; the field, in particular its horizontal component, is stronger in the 
chromopshere outside the spot and this is suggestive of a magnetic canopy. The sunspot displays a twist that 
on average is similar in the two layers. However, the differential twist between the photosphere and chromosphere
increases rapidly toward the outer penumbral boundary. The magnetic field vector is more horizontal with respect to the solar surface 
by roughly 5-20$^\circ$ in the photosphere compared to the upper chromosphere. Above a lightbridge, the chromospheric magnetic field
is equally strong as that in the umbra, whereas the field of the lightbridge is weaker than its surroundings in the photosphere 
by roughly 1\,kG. This suggests a cusp-like magnetic field structure above the lightbridge.}
   {}

\keywords{Sun: magnetic field - Sun: activity - Sun: chromosphere - Polarimetry} 

\maketitle
%

\section{Introduction} \label{sec_1}

The photospheric structure of the magnetic field of sunspots has been studied 
very extensively in the last few decades through Zeeman diagnostics from 
various magnetically sensitive spectral lines \citep{Solanki_2003, Borrero_2011}.
The study 
of the three-dimensional structure of sunspots up to the
chromosphere is much more challenging  observationally and therefore less explored.

The \ionn{He}{i} triplet at 10830\,\AA\, provides a promising avenue to study the upper 
chromospheric magnetic field \citep[see overview by][]{Lagg_2007,Lagg_2015}. The special 
formation process of this triplet \citep{Penn_1995,Rueedi_1995b} makes it a 
simpler tool to diagnose the magnetic field, compared to most other chromospheric spectral lines. 
Many studies of the magnetic field vector in the upper 
chromospheric layer use the \ionn{He}{i}\, 
triplet \citep[e.g.,][]{Rueedi_1996,Solanki_2003b,Lagg_2004,Trujillo_2005, Solanki_2006, 
Xu_2010,Merenda_2011,Xu_2012,Schad_2015,Joshi_2016}. The \ionn{He}{i}\, triplet is produced by transitions
between the $1s2s~^{3}S_{1}$ and the $1s2p~^{3}P_{0,1,2}$ energy levels. Extreme ultraviolet (EUV) radiation 
from the corona ionizes neutral helium atoms in the upper chromosphere, which then populate 
the lower level of the transition by recombination 
\citep{Avrett_1994,Andretta_1997,Centeno_2008}. The fine scale structures observed in the \ionn{He}{i}\,
triplet images are caused by ionizing radiation from the transition region \citep{Leenaarts_2016}.   

\citet{Trujillo_2002} have shown that the \ionn{He}{i} triplet observed in polarized light
is influenced by both the Zeeman effect and Hanle effect. At solar
disk center, the Hanle effect acts in forward scattering and can produce linearly polarized 
light in the \ionn{He}{i} triplet only in the presence of a magnetic field inclined with 
respect to the solar radius vector. In the presence of strong magnetic fields, as in sunspots,
linear polarization is dominated by the transverse Zeeman effect.

\citet{Rueedi_1995b}, \citet{Orozco_2005}, and \citet{Schad_2015} studied the upper chromospheric magnetic field of a sunspot
using the \ionn{He}{i} triplet and compared it with its photospheric 
counterpart. The vertical gradient of the magnetic field strength in the umbra found by 
\citet{Rueedi_1995b} is around 0.35-0.60\,G\,km$^{\rm{-1}}$ with positive values that denote 
increasing field strength with increasing depth. This value is similar to
that found by \citet{Abdussamatov_1971} who compared magnetograms derived from the H$\alpha$ line
with those derived from the \ionn{Fe}{i}\,6302.5\,\AA\, line. Values of the vertical magnetic field 
gradient found by \citet{Henze_1982} and \citet{Hagyard_1983}  fall in the same range.
They derived this value from the \ionn{C}{iv}\,1548\,\AA\, emission line in the transition region and 
the photospheric \ionn{Fe}{i}\,5250\,\AA\, absorption line. In penumbrae, \citet{Rueedi_1995b} 
reported a value of the vertical gradient of the magnetic field of around 
0.1-0.3\,G\,km$^{\rm{-1}}$. They also find a canopy-like structure in the longitudinal magnetic field
at the upper chromosphere around the sunspot \citep[see also][]{Schad_2015}. 
Recently, \citet{Joshi_2016} observed a spine and inter-spine structure in the magnetic field inclination of the penumbra
using high spatial resolution observations in the \ionn{He}{i}\, triplet obtained with the GREGOR Infrared Spectrograph
\citep[GRIS;][]{2012AN....333..872C} at the 1.5-meter GREGOR telescope \citep{Schmidt_2012}. 

Here we present inversions of the full Stokes vectors of the \ionn{He}{i}\, 
triplet at 10830\,\AA, the \ionn{Si}{i}\,10827.1\,\AA,\, and the \ionn{Ca}{i}\,10833.4\,\AA\
lines. The latter two  give us maps of the magnetic vector in  two different height layers of the photosphere. These maps are analyzed 
to discuss the differences between the photospheric and upper chromospheric magnetic 
field structure of the sunspot.

\begin{figure}
\centering
      \includegraphics[width = 0.45\textwidth]{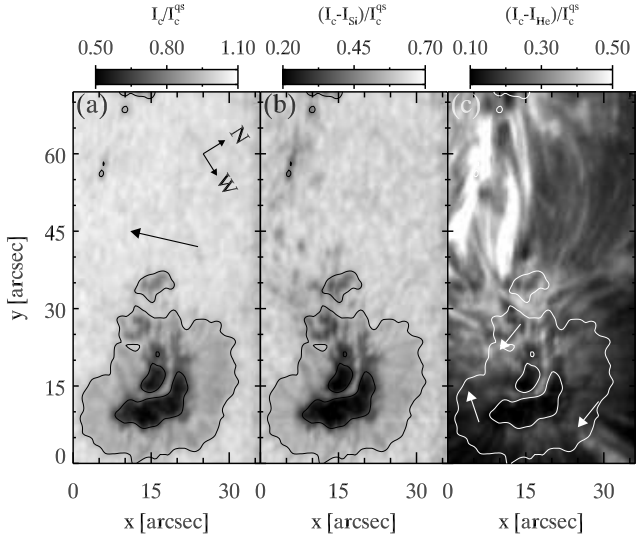}
      \caption{Panel (a) shows the observed field-of-view (FOV) on 14 November
       2010 in the continuum. Panels (b) and (c) depict depression of the 
       line core of the \ionn{Si}{i}\,10827\,\AA\, line and  \ionn{He}{i}bc, 
       respectively. The inner and outer contours 
       in all panels indicate the umbra-penumbra boundary and the outer boundary of the sunspot,
       respectively. The arrow in panel (a) indicates the direction to the solar disk center.
       Three white arrows in panel (c) mark positions of downflow 
        intrusions  in the penumbra in the upper chromosphere (see Fig.~\ref{maps14}(d)).}
\label{cont_14}
\end{figure}

\begin{figure}
\centering
      \includegraphics[width = 0.45\textwidth]{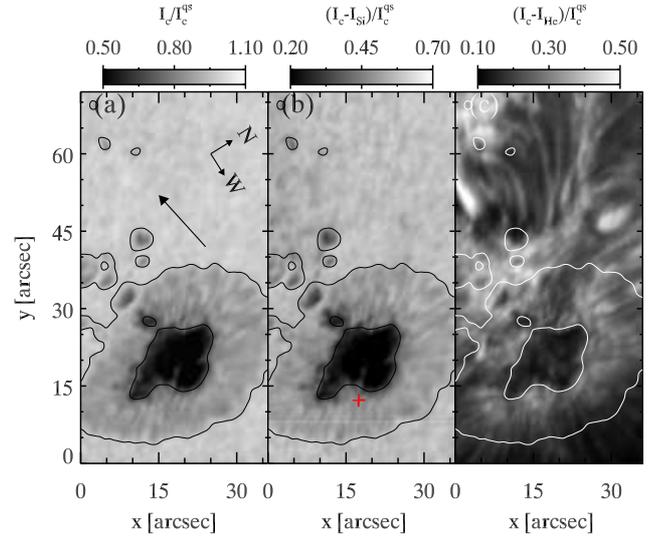}
      \caption{Same as Fig.~\ref{cont_14}, but showing the observed 
      FOV on 16 November 2010. The red plus in panel (b)
      indicates the position for which observed and the best fit 
      Stokes profiles are shown in Fig.~\ref{fit_helix}.}
\label{cont_16}
\end{figure}

\section{VTT/TIP-2 Observations} \label{sec_2}

We used the same data set as presented by \citet{Joshi_2017} (hereafter \citetalias{Joshi_2017}). 
The full Stokes vector of the photospheric
\ionn{Si}{i}\,10827.1\,\AA\, and \ionn{Ca}{i}\,10833.4\,\AA\, lines and the
upper chromospheric \ionn{He}{i} triplet at 10830\,\AA\, was recorded in a sunspot 
and its close surroundings using the Tenerife Infrared Polarimeter-2 
\citep[TIP-2;][]{Collados_2007} mounted on the German Vacuum Tower Telescope (VTT). 
The sunspot was recorded on 14 November 2010 (12\textdegree N, 10\textdegree W, $\mu$ = 0.96) 
and 16 November 2010 (14\textdegree N, 32\textdegree W, $\mu$ = 0.84). Atomic parameters of
the \ionn{He}{i} triplet are provided in Table~\ref{tab6.1}. Hereafter we refer to
the blue component of the triplet as \ionn{He}{i}a and the red components of the triplet as \ionn{He}{i}b and 
\ionn{He}{i}c. Since the latter two lines are blended, we refer to them together as \ionn{He}{i}bc.

\begin{table*}
\caption{Line parameters of the \ionn{He}{i} triplet at 10830\,\AA.\label{tab6.1}}
\begin{center}
\begin{tabular}{c c c c c c}
\hline\hline
Line & wavelength [\AA] & transition & $g_{\rm{eff}}$ & relative strength\\
\hline
\ionn{He}{i}a & 10829.09 & $1s2s~^{3}S_{1}$-$1s2p~^{3}P_{0}$ & 2.0     & 0.09\\
\ionn{He}{i}b & 10830.25 & $1s2s~^{3}S_{1}$-$1s2p~^{3}P_{1}$ & 1.75    & 0.30\\
\ionn{He}{i}c & 10830.34 & $1s2s~^{3}S_{1}$-$1s2p~^{3}P_{2}$ & 0.875   & 0.60\\
\hline

\end{tabular}
\end{center}
\end{table*}

Maps of the continuum intensity at 10832.6\,\AA, the depression of the \ionn{Si}{i}\, line core, 
$(I_{\rm{c}}-I_{\rm{Si}})/I_{\rm{c}}^{\rm{qs}}$, and 
of the line core of the \ionn{He}{i}bc, $(I_{\rm{c}}-I_{\rm{He}})/I_{\rm{c}}^{\rm{qs}}$,
observed on 14 November 2010 are shown in Fig.~\ref{cont_14}.
Here $I_{\rm{c}}$ and $I_{\rm{c}}^{\rm{qs}}$ represent the continuum intensity of the individual Stokes profile and the averaged continuum 
intensity in the quiet Sun, respectively. The intensities of the \ionn{Si}{i}\, line core and 
the \ionn{He}{i}bc\, line core are represented by $I_{\rm{Si}}$ and $I_{\rm{He}}$, respectively.
An arc filament can be seen in the \ionn{He}{i}bc line core depression map at position $x =18\varcsec - 26\varcsec$, 
$y = 5\varcsec - 38\varcsec$. Fig.~\ref{cont_16} shows maps of the continuum intensity at 10832.6\,\AA\, and the depression of the
\ionn{Si}{i}\, and \ionn{He}{i}bc line core 
observed on 16 November 2010.
The sunspot was observed in a growing phase on 14 November 2010; the increase of
its projected area from 450\,Mm$^2$\, on 14 November 2010 to 1054\,Mm$^2$\, on 16 
November 2010 corresponded to an increase of 234\%. If we correct for foreshortening then it 
grew by 281\% in two days.

\section{Inversions} \label{sec_2.1}

The \ionn{He}{i} absorption is thought to arise in a thin slab located at the upper boundary of the chromosphere.
The reason for this is the special formation process of the triplet requiring coronal UV illumination. 
The chromosphere is highly opaque to this radiation, which therefore affects only its uppermost layer, justifying the assumption that the atmospheric parameters do not vary within the thin slab where the \ionn{He}{i} triplet
forms. Hence, we inverted all three components of the \ionn{He}{i} triplet with the HeLIx$^{\rm{+}}$
inversion code assuming a Milne-Eddington type atmosphere. For details about 
HeLIx$^{\rm{+}}$, see \citet{Lagg_2004,Lagg_2009}.  
The HeLIx$^{\rm{+}}$ code includes a consideration of the incomplete Paschen-Back effect
regimes \citep[cf.][]{Socas_2005b, Sasso_2006}. 

\begin{figure}
\centering
      \includegraphics[width = 0.45\textwidth]{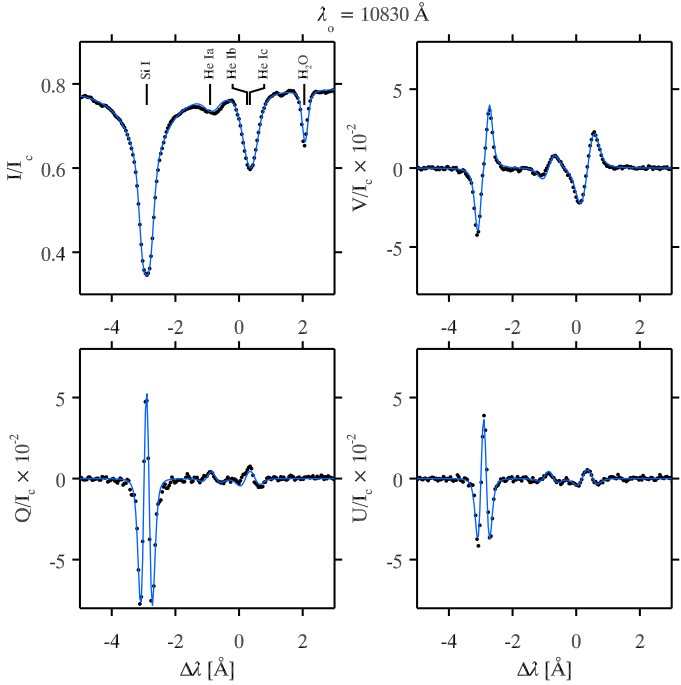}
      \caption{Best fit HeLIx$^{\rm{+}}$ inversions of typical Stokes profiles
               in the penumbra. Their spatial location is marked by a red plus 
               sign in Fig.~\ref{cont_16}(a). 
               Black dots represent observed data points
               and solid blue curves represent the best fits.}
\label{fit_helix}
\end{figure}

The blue component of the \ionn{He}{i} triplet is blended by the red wing of the \ionn{Si}{i}
line. This blending is taken into account in a self-consistent manner by inverting 
the \ionn{He}{i} triplet and \ionn{Si}{i} line simultaneously. Our model atmosphere used for this purpose consists of 
eight free parameters to fit the observed Stokes profiles of the \ionn{He}{i} triplet: the magnetic field strength, $B$, the inclination of 
the magnetic field vector, $\gamma$, the azimuth angle of the magnetic field vector, $\phi$, 
the LOS velocity, $v_{\rm{los}}$,  the Doppler width, $\Delta\lambda_{\rm{D}}$, the damping
constant, $a$, the gradient of the source function, $S_{\rm{1}}$, and the opacity
ratio between line center and continuum, $\eta_{\rm{0}}$.  
To fit the \ionn{Si}{i} line, the model atmosphere consists of the same free parameters as the 
\ionn{He}{i} triplet along with two additional free parameters, 
the global stray-light factor, $\alpha$ and the line-of-sight velocity for the global  
stray-light components. The approach for global stray light is similar to 
that used in \citetalias{Joshi_2017}, where the global stray light is assumed to originate from the broad wings
of the point-spread function and therefore resembles the shape of the
average quiet Sun Stokes $I$ profile.

\begin{figure*}
\centering
      \includegraphics[width = 1.0\textwidth]{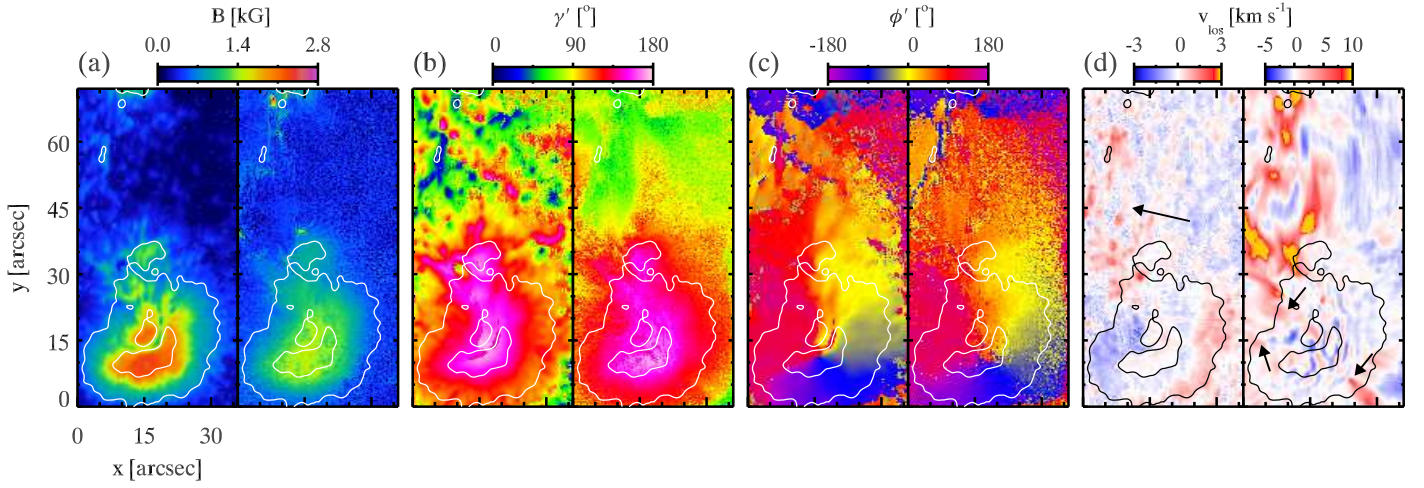}
      \caption{Maps of (a) Magnetic field strength $B$,
               (b) inclination angle, $\gamma^{\prime}$, of the magnetic field vector
               with respect to the solar surface normal, 
               (c) azimuth direction, $\phi^{\prime}$, of the magnetic field vector and 
               (d) line-of-sight velocity obtained from the observations 
               recorded on 14 November 2010. Left and right maps in all panels represent 
               atmospheric parameters obtained in the photosphere and upper
               chromosphere, respectively. An arrow in the left map at of  
               panel (d) indicates the disk center direction and three arrows in the right map  
               point three downflow intrusions in the penumbra.}
\label{maps14}
\end{figure*}

\begin{figure*}
\centering
      \includegraphics[width = 1.0\textwidth]{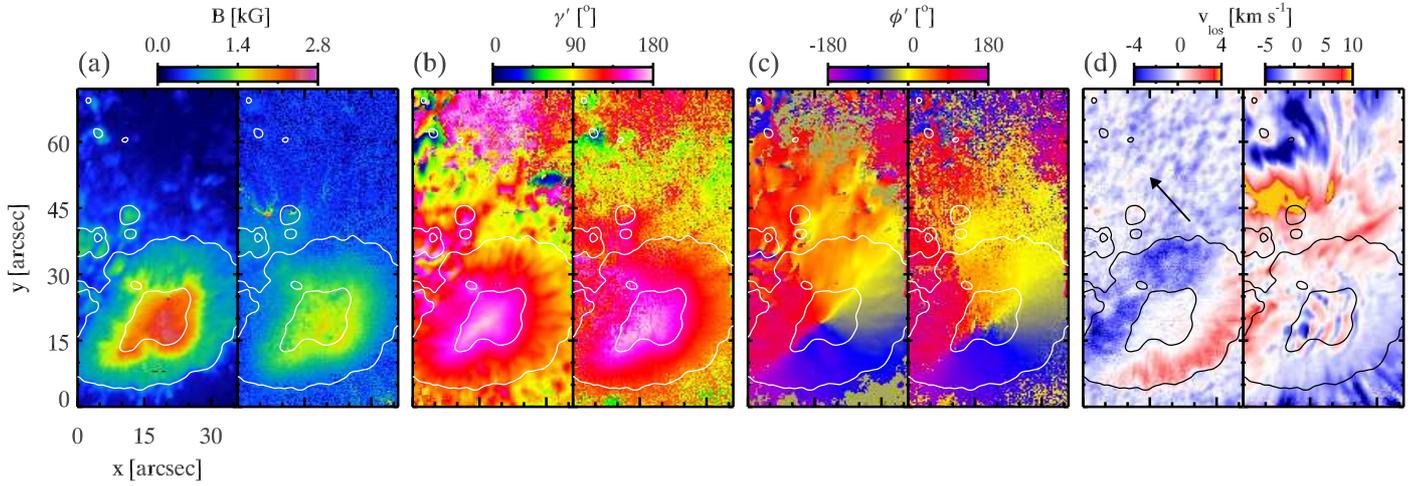}
      \caption{Same as Fig.~\ref{maps14}, but for the observations
               recorded on 16 November 2010.}
      \label{maps16}
\end{figure*}

To take into account the blends from the \ionn{Si}{i}\, line into the \ionn{He}{i}\,  
triplet in the inversions, we took a multi-step approach. We first inverted the Stokes profiles from the \ionn{Si}{i}\, line.
Then, in a second run, we fitted the Stokes profiles of the \ionn{He}{i}\, triplet and we fixed
the free parameters for the \ionn{Si}{i} line to the values retrieved from the first run. Finally, we
fitted both the \ionn{Si}{i} line and \ionn{He}{i} triplet simultaneously, but we allowed the 
parameters to vary only within $\pm5\%$ of the values fitted in the previous runs. The $\rm{H_{2}0}$
telluric line at 10832\,\AA, which can blend the \ionn{He}{i}\, triplet when it is strongly redshifted, is fitted with a Voigt function.
Observed Stokes profiles at one penumbral pixel along with the best fits are depicted in Fig.~\ref{fit_helix}.
The purpose of including the \ionn{Si}{i}\, line in the inversions of the \ionn{He}{i}\, triplet is
only to account for blending of the \ionn{He}{i}\, triplet. 

For the analysis presented in this paper we use the magnetic field vector in the
photosphere inferred from the SPINOR \citep{Frutiger_1999,Frutiger_2000} inversions of the \ion{Si}{i}\, and
\ionn{Ca}{i}\, lines. The combination of the strong \ionn{Si}{i}\, line and weak \ionn{Ca}{i}\, line
puts more constraints especially on the height information of the free parameters in the inversion process compared to an  
inversion of a single line only. Our model atmosphere consists of three nodes, 
$\log \tau_{\rm{630}} = 0.0, -0.7,$ and $-2.3$ for the line-of-sight (LOS) velocity, 
$v_{\rm{los}}$, and the temperature, $T$, where $\tau_{\rm{630}}$ corresponds to the optical depth at 630\,nm.
The magnetic field strength is assumed to vary linearly with respect to $\log \tau$.
The other atmospheric parameters, such as inclination of the magnetic field relative to 
LOS, $\gamma$; its azimuth direction, $\phi$; and the micro-turbulent velocity, 
$v_{\rm{mic}}$, were assumed to be constant with height. Only the height-dependent atmosphere inferred from
the SPINOR inversions can take into account the strong asymmetries due
to gradients in the Doppler velocities and the magnetic field in the Stokes profiles of the \ionn{Si}{i}\, line.
More detailed information on the inversions are mentioned in \citetalias{Joshi_2017}.

\section{Analysis and results} \label{sec_3}

Maps of the magnetic field vector and the LOS velocity retrieved from the observations 
recorded on 14 November 2010 are shown in Fig.~\ref{maps14}.
In each panel the maps on the left correspond to the parameters obtained in the photosphere 
through the SPINOR inversions of the \ionn{Si}{i}\, and \ionn{Ca}{i}\,
lines. In the SPINOR inversions we obtained $B$ and its linear gradient with respect to 
$\log \tau$, but for comparison of the photospheric magnetic field properties with
the upper chromosphere we use values of $B$ obtained by averaging between $\log \tau$  =  0.0 and 
$\log \tau$  =  --2.3. The atmospheric parameters obtained from the 
upper chromosphere through the HeLIx$^{\rm{+}}$ inversions of the \ionn{He}{i}\, triplet 
are shown in the right panels. Fig.~\ref{maps16} shows the same plots for 
the data observed on 16 November 2010. 

Overall, within the visible 
boundary of the sunspot the magnetic field strength in the upper chromosphere is weaker than in the photosphere. 
This is particularly striking in the umbra, as can be seen 
in the maps of $B$ from both days. The lightbridge observed on 14 November 2010 shows 
weaker $B$ compared to the umbra in its photospheric layer and in the upper chromosphere no
signature of the lightbridge is present in the $B$ map. The values of $B$ in the 
upper chromosphere at the location of the lightbridge are comparable to the umbral 
magnetic field strength, although the \ionn{He}{i} line depth in the lightbridge is 
considerably bigger than in the surrounding umbra (see Fig.~\ref{cont_14}).

The 180\textdegree\, ambiguity in the azimuth direction was resolved by applying the ``acute angle'',
method \citep{Sakurai_1985,Cuperman_1992}. The magnetic field 
vectors presented here are projected to disk center coordinates using the transformation 
matrix by \citet{Wilkinson_1989}. The inclination angle with respect to solar surface normal and
the ambiguity resolved azimuth angle are denoted by $\gamma^{\prime}$ and $\phi^{\prime}$,
respectively.

\begin{figure*}
\centering
      \includegraphics[width = 1.0\textwidth]{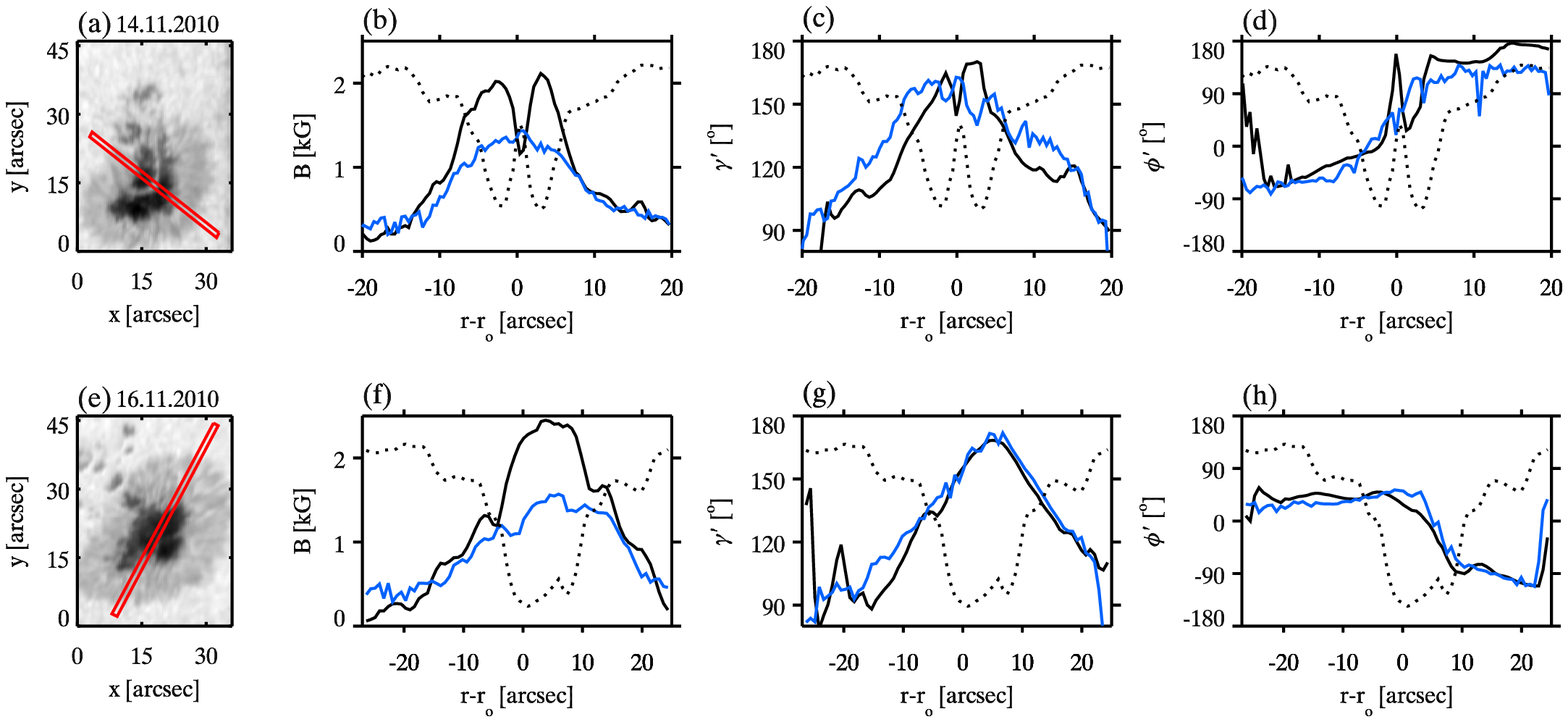}
      \caption{Cross sections through the observed sunspots: 
               Panels (b) and (f), (c) and (g) and (d) and (h) depict profiles of $B$, $\gamma^{\prime}$ and
               $\phi^{\prime}$, respectively, along an artificial slit represented by the 
               red lines in panels (a) and (e). Panels (a), (b), (c), and (d)
               refer to observations recorded on 14 November 2010 and (e), (f), (g), and (h) correspond to
               16 November 2010.  The black and blue curves in panels (b), (c), (d), (f)
               (g), and (h) correspond to the photosphere and upper chromosphere,
               respectively, while the dotted curves represent the continuum intensity.}
\label{slit_14_16}
\end{figure*}

The inclination of the magnetic field vector in the upper chromosphere
looks qualitatively similar to that in the photosphere. In the photosphere, 
outside the boundary, but close to the sunspot, we see a few small patches where $\gamma^{\prime}$ has values 
less than 90\textdegree, i.e., at those locations the magnetic field has polarity that is opposite to
that of the umbra. These  opposite polarity patches are not visible in the upper chromosphere, suggesting 
that they do not reach the height of the formation of the \ionn{He}{i}\, triplet, as it is 
covered by the canopy of the sunspot. Maps of $\gamma^{\prime}$ from both days
show more fine structure in the photosphere as compared to the upper chromosphere.
In the sunspot the magnetic field azimuth in the upper chromosphere
is generally similar to that in the photosphere. One exception is the lightbridge, 
which shows an indication of the field from the umbra expanding over 
it in the photosphere on both sides, but no signature at all in the upper chromosphere. Outside the sunspot, 
maps of the magnetic field vectors in the upper chromosphere are much noisier than in
the photosphere. This has its source in the much weaker Stokes $Q$ and $U$ in the \ionn{He}{i}\,
triplet compared to the \ionn{Si}{i}\, line. 

The LOS velocity maps are shown in Fig.~\ref{maps14}(d) and Fig.~\ref{maps16}(d). 
In the photosphere these refer to $\log \tau$ = 0.0. We use this layer since it 
displays stronger signature of the Evershed flow than the other
nodes higher in the atmosphere. 
The magnitude of velocities is considerably stronger on 16 November 2010, when the sunspot was closer to the limb ($\mu$ = 0.84).
The LOS velocities derived in the upper chromosphere show radial inflows on 
16 November 2010, which is consistent with the inverse Evershed effect. On 14 November 2010 the
situation is less clear cut with no clear sign of the inverse Evershed effect in the \ionn{He}{i}
triplet. Three intrusions of denser (brighter in Fig.~\ref{cont_14}(c)) 
downflowing gas are identified in the chromosphere (see small arrows in Fig.~\ref{cont_14}(c) and Fig.\ref{maps14}(d)).  
These may be associated with sunspot plumes \citep{Fludra_1997,Maltby_1998,Maltby_1999,Brynildsen_1998,
Brynildsen_1999,Brynildsen_2001,Fludra_2001,Brosius_2004,Brosius_2005}.
The maps of $v_{\rm{los}}$ derived from the \ionn{He}{i} triplet show wave-like 
structures in the sunspot umbra and in the inner penumbra. This wave structure is 
elongated in the direction of the $y$-axes (i.e., in the direction of the slit of the spectrograph). 
This pattern is produced by running penumbral waves and umbral flashes \citep{Zirin_1972,
Christopoulou_2000,Georgakilas_2000,Christopoulou_2001,Bogdan_2006,
Centeno_2006,Tziotziou_2006,Tziotziou_2007, Bloomfield_2007b,Felipe_2010,de_la_cruz_roderiguez_2013}.

We expect a small offset between the parameter maps obtained in the photosphere and upper chromosphere
because of the viewing geometry; the sunspot was observed away from disk center on both days.
If we assume a height difference of 1000\,km \citep[see][]{Centeno_2006} between the two 
observed layers of the atmosphere, which is a conservative estimate, then the chromospheric parameter maps are offset by $\sim$275\,km and $\sim$575\,km toward
the limb on 14 and 16 November 2010, respectively. The analysis described in the following sections was
carried out after correcting for these offsets.

For a more detailed and quantitative insight into the magnetic field properties of
the sunspot and the connection between its photospheric layer and upper chromospheric layer,
$B$, $\gamma^{\prime}$, and $\phi^{\prime}$ are plotted along an artificial $3.0\varcsec$ wide slit laid across the sunspot 
(see Fig.~\ref{slit_14_16}). 
The parameters $B$, $\gamma^{\prime}$, and $\phi^{\prime}$ are averaged perpendicularly to the slit direction. 
On 14 November 2010 the profile of $B$ along the slit in the photosphere shows a dip at 
the location of the lightbridge with a minimum value of $B$ $\sim$1.2~\,kG,\, 
which is $\sim$1.0~\,kG\, weaker than in the surrounding umbra. In the upper chromosphere,
at the location of the lightbridge $B$ is $\sim$1.4~\,kG. This value is comparable with and 
maybe even slightly higher than the value in the surrounding umbra and higher than in the
photopshere. A similar $B$ in a lightbridge and umbra in chromospheres  was already noticed by
\citet{Rueedi_1995b}.
The difference between the photospheric
and upper chromospheric magnetic field strength becomes rather small in the penumbra.
Just outside of the visible boundary of the sunspot, the values of $B$ are higher by up to $\sim$300\,G\, in the 
upper chromosphere as compared to the photosphere. \citet{Rueedi_1995b} interpreted
this behavior as evidence of a magnetic canopy.

Profiles of $\gamma^{\prime}$ along the slit on 14 November 2010 indicate that inside the sunspot,
except in the inner part of the umbra, the magnetic field is in general more horizontal in the photosphere
compared to the upper chromosphere. This is particularly pronounced in the penumbra and 
more so on the 14 November 2010. At the center of the lightbridge the magnetic field 
in the photosphere becomes more horizontal in agreement with results in the literature \citep[see, e.g.,][and references therein]{Lagg_2014}.
In the chromosphere the field in the center of the lightbridge is as vertical as everywhere else
along the cut and more vertical than in the umbra directly at the edge of the lightbridge. In summary, the field in the light bridge is 
more vertical in the upper chromosphere than in the photosphere. The umbra close to 
the lightbridge shows the opposite trend. On 16 November 2010, the photospheric and the 
upper chromospheric profile of $\gamma^{\prime}$ are quantitatively more similar.

On 14 November 2010 profiles of $\phi^{\prime}$ in the photosphere along the slit show a 
sudden change of $\sim$180$^\circ$ (compared to that in the nearby umbra) at the both edges of the lightbridge, which suggests that the magnetic field lines are  
fanning out at the the location of the lightbridge. The sunspot has negative polarity and so magnetic field line points inward.
Variation in  $\phi^{\prime}$ in the upper chromosphere is rather steady.

The small difference in the magnetic field strength and inclination between the lightbridge and its immediate surroundings may be due to 
a difference in the formation height of the \ionn{He}{i}\,10830\,\AA\, over the lightbridge 
compared to the umbra. The sudden difference in the gradient of $B$ between umbra, on the one hand, 
and the penumbra and lightbridge, on the other, also suggest a difference in formation height.

\begin{figure}[b]
\centering
      \includegraphics[width = 0.49\textwidth]{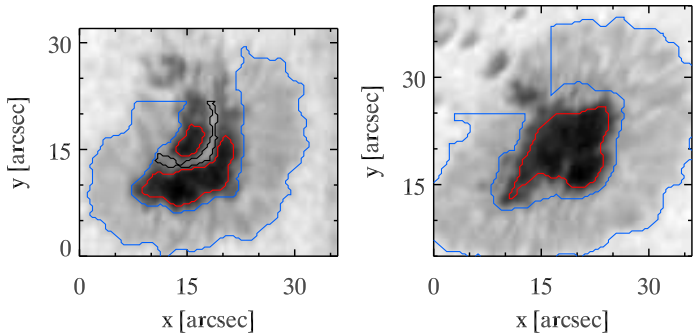}
      \caption{Division of the  sunspot into umbral, penumbral, and lightbridge areas. Red and blue contours in 
      both panels encircle umbral and penumbral areas, which are used to compare the magnetic field vector 
      retrieved in the photosphere and upper chromosphere. The area within the black
      contour in the left panel represents the lightbridge. The left and right
      panels correspond to the observations recorded on 14 and 16 November 2010, respectively.}
\label{class_14_16}
\end{figure}

\begin{figure}
\centering
      \includegraphics[width = .45\textwidth]{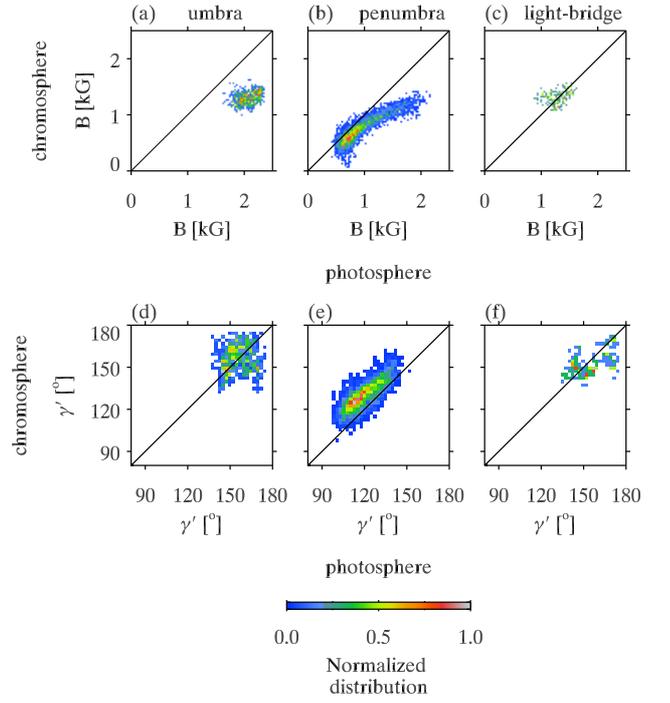}
      \caption{Two-dimensional histograms of the magnetic field components obtained 
               in the upper chromosphere vs. those obtained in the photosphere 
               from the observations recorded on 14 November 2010.
               Panels (a), (b), and (c) show plots of $B$ from
               the umbral, penumbral , and lightbridge pixels,
               respectively (see Fig.~\ref{class_14_16}).\ The $\gamma^{\prime}$ values of data points from the same spatial 
               positions are shown in panels (d), (e), and (f),
               respectively.}
                            
\label{sct_14}
\end{figure}

\begin{figure}
\centering
      \includegraphics[width = 0.30\textwidth]{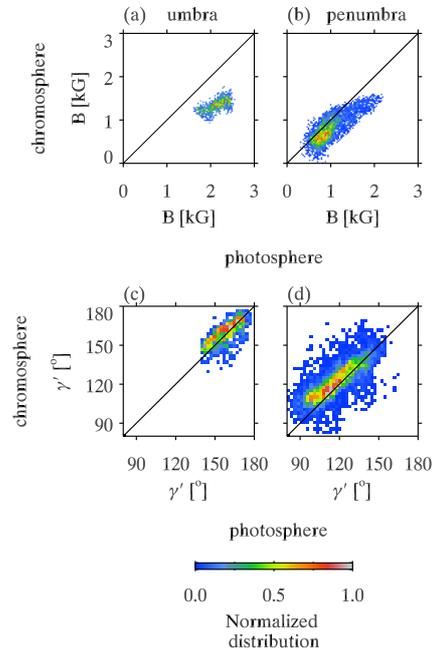}
      \caption{Same as Fig.~\ref{sct_14} 
                but for the observations recorded on 16 November 2010 and 
                restricted to umbral and penumbral pixels.}
\label{sct_16}
\end{figure}

To emphasize the differences between the umbra, penumbra, and lightbridge,
we show two-dimensional histograms of the magnetic field components in the upper chromosphere versus
those of the photosphere in Fig.~\ref{sct_14} and Fig.~\ref{sct_16} 
for 14 and 16 November 2010, respectively. The areas considered here for the different 
regions of the sunspot are indicated in Fig.~\ref{class_14_16} with red, 
blue and black contours. The blue contours exclude the strongly distorted parts of the penumbra. Also,
the boundary region between the umbra and penumbra is excluded to make sure that the pixels under consideration
are either from the umbra or penumbra and to the avoid the transition between them in the analysis 
as such pixels are likely to be more affected by stray light.   

All the pixels in the umbrae indicate that the $B$ in the upper chromosphere 
is lower by a factor of 1.30-1.65 than that in the photosphere on both days of the observations. 
In the penumbra the magnetic field strength in the upper chromosphere is,
in general, a factor 1.00-1.55 lower than that in the photosphere, while a few pixels even indicate 
higher values in the upper chromosphere.
The lower the penumbral field strength, $B$, the smaller the difference between the photosphere and 
upper chromosphere.

The two-dimensional histograms of $\gamma^{\prime}$ from the observations on both days 
indicate that the upper chromospheric magnetic field is more vertical compared 
to the photospheric magnetic field by roughly 5-20\textdegree. The signal-to-noise ratio
is lower in the \ionn{He}{i} triplet, especially for $Q$ and $U$. Since low signal-to-noise ratios 
lead inversion codes to return to more horizontal \textit{\textbf{B}} 
\citep{Borrero_2012,Jafarzadeh_2013,Jafarzadeh_2014}, this result is not an artifact of the noise. 
The values of $B$ and $\gamma^{\prime}$ in the lightbridge observed on 14 November 2010
are comparable both in the photosphere and in the upper chromosphere.

\begin{figure}
\centering
      \includegraphics[width = 0.48\textwidth]{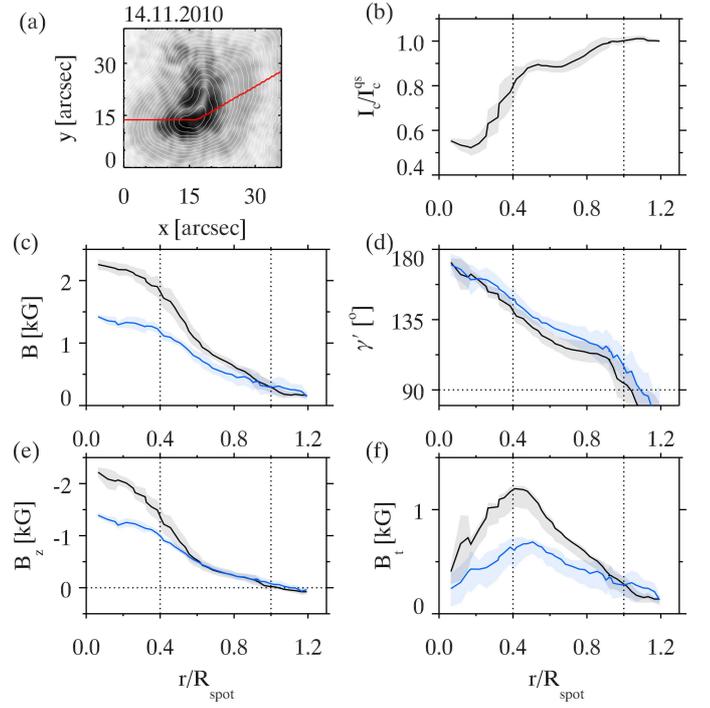}
      \caption{Averaged radial distribution of the magnetic field properties in the 
      observed sunspot: Panel (a) depicts the continuum intensity map with overplotted contours 
      used to calculate azimuthal averages for 14 November 2010.
      The area above the red line was not included to calculate the azimuthal averages. 
      Panel (b) shows the relative continuum intensity (normalized to the quiet Sun) 
      of the sunspot as a function of the normalized radial distance, $r/R_{\rm{spot}}$.
      Panels (c), (d) (e), and (f) depict the radial dependence of $B$, $\gamma^{\prime}$,
      the vertical component of the magnetic field, $B_{\rm{z}}$, and the transverse 
      component of the magnetic field, $B_{\rm{t}}$, respectively. The
      black and blue curves correspond to the magnetic field vector derived 
      in the photosphere and upper chromosphere, respectively. 
      The dotted vertical lines in panels (b)-(f) indicate
      the umbra-penumbra boundary and the outer boundary of the sunspot.}
\label{radial_14}
\end{figure}

\begin{figure}
\centering
      \includegraphics[width = 0.48\textwidth]{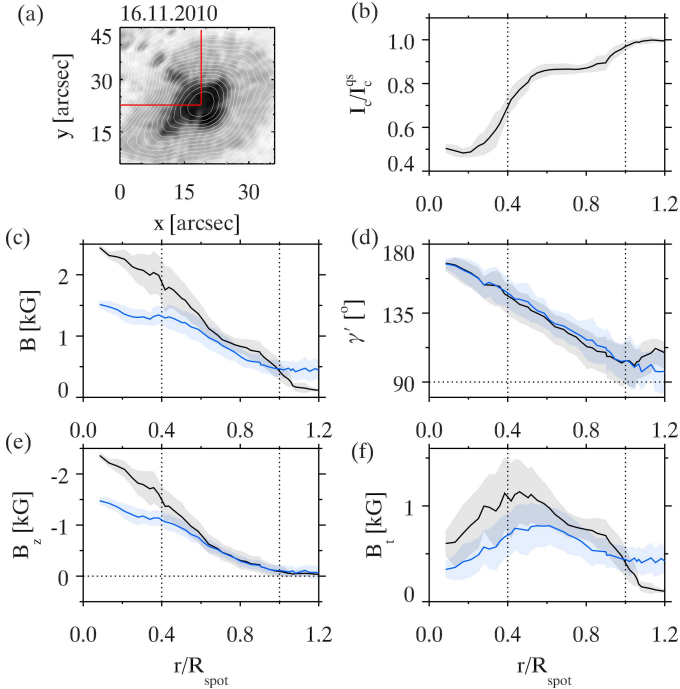}
      \caption{Same as Fig.~\ref{radial_14}, but for
      the observations recorded on 16 November 2010. 
      Here the area to the upper left of the red
      lines is not included in making the radial averages.}
\label{radial_16}
\end{figure}

We take azimuthal averages along smoothed iso-intensity contours to derive the average radial dependence of the components of the magnetic
field vector.
Figs.~\ref{radial_14}(a) and~\ref{radial_16}(a) show these contour lines. 
Areas above the red line are excluded in the computation of azimuthal averages
because of their complex photospheric structure. The remainder of the panels depicts 
the azimuthal averages of $I_{c}$ and various magnetic parameters and corresponding 
standard deviations as a function of normalized radial distance, 
$r/R_{\rm{spot}}$, from the sunspot center for 14 and 16 November 2010.

On 14 November\ 2010, $B$ has an average value of
$\sim$2.3~\,kG\, in the darkest part of the umbra in the photosphere, decreasing to $\sim$1.8~\,kG\,
at the umbra-penumbra boundary. In the upper chromosphere $B$ amounts to 
$\sim$1.3~\,kG\, at the sunspot center and it remains almost constant 
to the umbra-penumbra boundary. On 16 November
2010 the center of the sunspot has a value of $B$ around $\sim$2.5~\,kG\, and of $\sim$1.9~\,kG\ 
at the umbra-penumbra boundary in the 
photosphere. In the upper chromosphere $B$ is nearly constant at a value 
$\sim$1.4~\,kG\, from the sunspot center to the umbra-penumbra boundary.
The difference in the values of $B$ between the photosphere and upper chromosphere
decreases from the umbra-penumbra boundary toward the outer boundary of 
the sunspot; the same field strength of $\sim$400~\,G\, at both layers is reached close to 
the outer sunspot boundary. This is valid for both days. Outside the boundary of the sunspot, $B$ is larger in the upper chromosphere than in the photosphere on both days.

Azimuthally averaged  radial profiles of $\gamma^{\prime}$ from  both days suggest that
the magnetic field in the upper chromosphere is more vertical than
the magnetic field in the photosphere both in the umbra and the penumbra (see Fig.~\ref{radial_14} and \ref{radial_16}). 
The magnetic field in the upper chromosphere is on average 10-20\textdegree\,
less inclined compared to the photosphere on 14 November 2010. On
16 November 2010 it is only 5-10\textdegree\, less inclined in the  
upper chromosphere.

The vertical component of the magnetic field, $B_{\rm{z}}$, shows a negligible 
difference between the upper chromosphere and the photosphere for 
$r\geq$ 0.60$R_{\rm{spot}}$ on both days. The maximum of the transverse component 
of the magnetic field, $B_{\rm{t}}$, in the photosphere is found at the 
$r\simeq 0.40 - 0.45R_{\rm{spot}}$ while in the upper chromosphere it is at 
$r\simeq 0.50 - 0.60R_{\rm{spot}}$. This move of the maximum horizontal field 
to further out in the spot is consistent with the fact that the horizontal field 
is considerably larger in the upper chromosphere than in the photosphere outside 
the outer boundary of the spot. This is particularly evident on 16 November 2010. The chromospheric field is also more horizontal there. These properties 
are consistent with the presence of a magnetic canopy whose lower boundary may 
lie between the two layers considered here.

We estimated the vertical gradient of the magnetic field, $\Delta B/\Delta d$
(here $d$ denotes the geometrical depth, i.e., $d$ increases into the Sun), 
between the photosphere and upper chromosphere by assuming that the \ionn{He}{i} triplet forms approximately 1000\,km 
above the formation height of the \ionn{Si}{i}\, line in the sunspot. This difference is inferred from the
study of \citet{Centeno_2006} who retrieved this height difference 
between the photosphere and formation height of the \ionn{He}{i}\, triplet in the sunspot 
umbra by analyzing phase spectra of LOS velocities inferred from the
\ionn{Si}{i}\, and \ionn{He}{i}\, triplet.
We are aware that the formation height of the \ionn{He}{i} triplet could be 
significantly different for the umbra and penumbra.
\citet{Joshi_2016} estimated the difference in the formation height of these lines to be 1250\,km in sunspot penumbrae by looking at
the shift in the apparent neutral lines in the Stokes $V$ images of the respective spectral lines. The method takes into account the difference in the inclination 
of magnetic field at a certain radial position in the penumbra. 
According to Fig.~\ref{radial_14} the field in the chromosphere 
is somewhat more vertical, although Fig.~\ref{radial_16} shows that this difference can be quite small for a relatively regular spot. 
This implies that the 1250 km found by \citet{Joshi_2016} is an upper limit for the height difference. In the following 
we therefore mainly employ the 1000 km given by \citet{Centeno_2006}.

We learned from \citetalias{Joshi_2017} that due to the highly corrugated iso-$\tau$ surfaces 
of sunspot penumbrae we can see a decreasing magnetic field strength within the photosphere with optical depth in the azimuthal averages. 
So, it is important to know how this configuration affects the estimation of $\Delta B/\Delta d$ between the 
photosphere and upper chromosphere. This is why we also estimate
$\Delta B/\Delta d$ between $\log \tau$ = 0.0 and the upper chromosphere and 
between $\log \tau$ =  --2.3 and the upper chromosphere. First we 
computed the difference of geometrical depth, $d_{\rm{diff}}$ between $\log \tau$ = 0.0 
and $\log \tau$ =  --2.3 for each pixel, assuming hydrostatic equilibrium. 
Then the geometrical depth difference, $\Delta d$, between $\log \tau$ = 0.0 and the upper chromosphere 
and between $\log \tau$ =  --2.3 and the upper chromosphere is estimated to be 
$1000 + (d_{\rm{diff}}/2)$\,km and $1000 - (d_{\rm{diff}}/2)$\,km, respectively.

\begin{figure}
\centering
      \includegraphics[width = 0.38\textwidth]{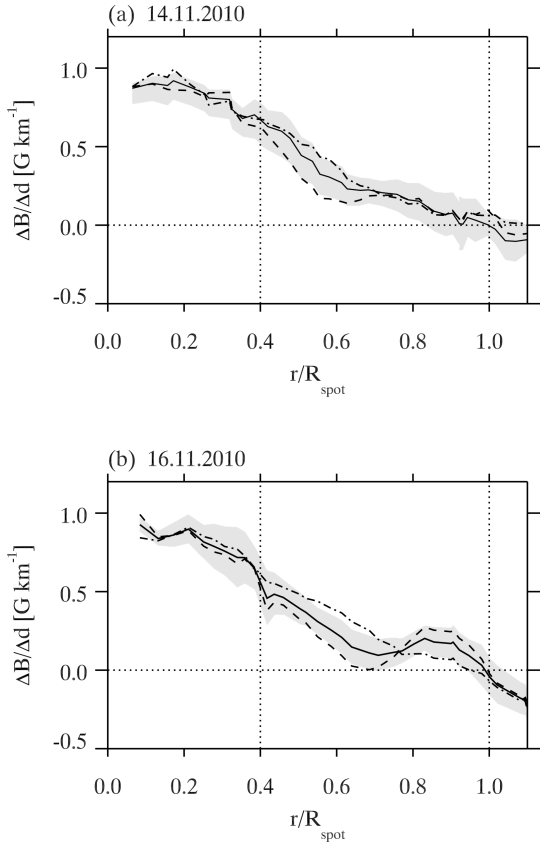}
      \caption{Vertical gradient of $B$, $\Delta B/\Delta d$, between
               the photosphere and upper chromosphere as a function of 
               $r/R_{\rm{spot}}$. Solid curves represent $\Delta B/\Delta d$
               estimated with the average value of $B$ in the photosphere.
               Dashed and dot-dashed curves show $\Delta B/\Delta d$ 
               from $\log \tau$ = 0.0 and $\log \tau$ =  --2.3, respectively.
               Panels (a) and (b) correspond to 14 and 16 November 2010,
               respectively.}     
\label{radial_grad}
\end{figure}

The radial dependence of $\Delta B/\Delta d$  between the photosphere 
and upper chromosphere is shown in Fig.~\ref{radial_grad} for both days.
Solid curves represent $\Delta B/\Delta d$ between the average photosphere 
and upper chromosphere. The value $\Delta B/\Delta d$ from $\log \tau$ = 0.0 and 
$\log \tau$ =  --2.3 to the upper chromosphere is represented by dot-dashed 
dashed curves, respectively. In the center of the umbra $\Delta B/\Delta d$ is similar for both $\log \tau$ layers,
irrespective of the photospheric layer used in estimating the gradient.
$B$ increases with depth in the darkest part of the umbra, i.e., on average by
0.9\,G\,km$^{-1}$ between the upper chromosphere and photosphere. From the center of the sunspot toward 
its outer boundary, the values of $\Delta B/\Delta d$ 
decrease, i.e., $B$ increases with geometrical 
depth with a slower rate in the penumbra compared to umbra.
The way $\Delta B/\Delta d$ drops with $r/R_{\rm{spot}}$ depends significantly
on the layer at which the photospheric field is taken. The higher this is the closer to
linear $\Delta B/\Delta d$ becomes with respect to $r/R_{\rm{spot}}$. As mentioned earlier, the difference in the formation height between the \ionn{He}{i}\, triplet and the \ionn{Si}{i}\,10827.1\,\AA\,
line can be significantly different from the umbra to that in the prnumbra; similarly, it may also change from one sunspot to another sunspot.
If we consider the difference in the formation heights 
to be 1250\,km,\, following \citet{Joshi_2016}, then the resulting vertical gradient of the magnetic field on average would be 
$\sim$0.7\,G\,km$^{-1}$ in the darkest part of the umbra. We consider this to be a lower limit to the gradient.

\begin{table*}
\begin{center}
\caption{Statistics of the twist and differential twist. ’A’ and ’B’ indicate two parts of the sunspot as indicated in Fig.~\ref{rot_14_16}, 
         ’All’ includes both parts. The azimuthal angles from the inversion, $\phi^{\prime}$, and from the potential field extrapolation, $\psi^{\prime}$,
         are specified for the chromosphere and photosphere (indices $_{\rm{Ch}}$ and $_{\rm{Ph}}$)}.\label{tab6.4}
\begin{tabular}{cccccc}
  \hline\hline
  \multicolumn{2}{c}{}                                            & \multicolumn{2}{c}{14 November 2010}        &  \multicolumn{2}{c}{16 November 2010}\\
  \hline
   Twist &  Area under consideration                              &   Mean [\textdegree] & FWHM [\textdegree]    & Mean [\textdegree]   & FWHM [\textdegree] \\     
  \hline 
                                                          & A     &   --16               & 40                    &  --24                & 30\\
  $\phi^{\prime}_{\rm{Ph}}$ -- $\psi^{}_{\rm{Ph}}$        & B     &   5                  & 20                    &  1                   & 30\\
                                                          & All   &   --6                & 30                    &  --13                & 50\\
  \hline
                                                          & A     &   --16               & 50                    &  --25                & 60\\
  $\phi^{\prime}_{\rm{Ch}}$ -- $\psi^{}_{\rm{Ch}}$        & B     &   4                  & 40                    &  4                   & 40\\
                                                          & All   &   --7                & 40                    &  --12                & 50\\
  \hline
                                                          & A     &   0                  & 30                    &  --1                 & 20\\
  $\phi^{\prime}_{\rm{Ph}}$ -- $\phi^{\prime}_{\rm{Ch}}$  & B     &   --1                & 40                    &  3                   & 20\\
                                                          & All   &   --1                & 30                    &  1                   & 20\\
  \hline
\label{table2}
\end{tabular}
\end{center}
\end{table*}

\begin{figure}
\centering
      \includegraphics[width = 0.45\textwidth]{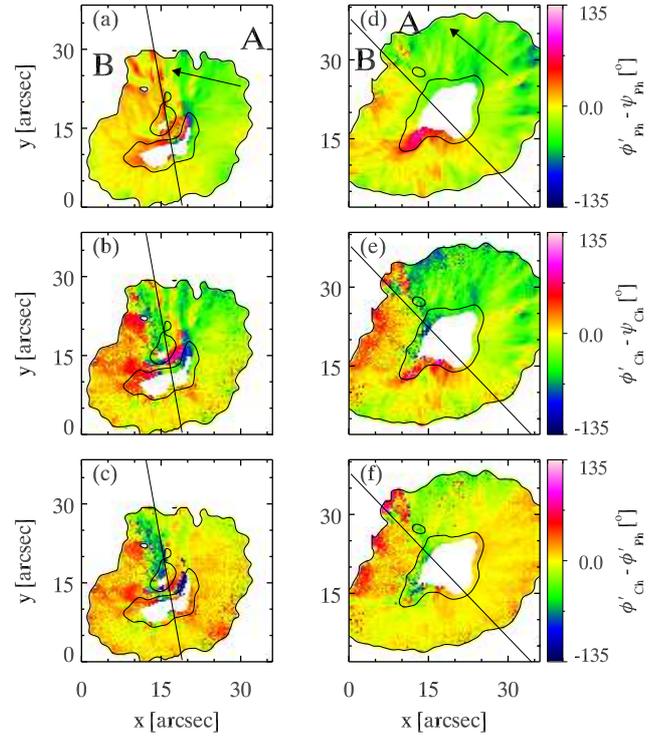}
      \caption{Maps of the twist and differential twist angles of magnetic field of the sunspot. 
               Panels (a) and (d): The twist of the photospheric field defined as the difference between the azimuth angle 
               of the photospheric field $\phi^{\prime}_{\rm{Ph}}$ and that of a potential field $\psi^{}_{\rm{Ph}}$, 
               $\phi^{\prime}_{\rm{Ph}}-\psi^{}_{\rm{Ph}}$. Panels (b) and (e): 
               Twist of the field at the chromospheric layers, $\phi^{\prime}_{\rm{Ch}}-\psi^{}_{\rm{Ch}}$,
               where $\phi^{\prime}_{\rm{Ch}}$ is the azimuth angle of the measured chromospheric field.
               Panels (c) and (f): Differential twist, i.e., the difference between the azimuth angle of
               the upper chromosphere and that of the photosphere, $\phi^{\prime}_{\rm{Ch}}-\phi^{\prime}_{\rm{Ph}}$.
               Panels in the left and right columns correspond to the observations recorded on
               14 November 2010 and 16 November 2010, respectively. Arrows in the panels
               (a) and (d) indicate the direction to disk center. On both days the sunspot
               is divided in two part denoted as 'A' and 'B' in panels (a) and (b). Statistics of the twist and differential twist 
               are presented in Fig.~\ref{hist} and Table~\ref{table2} for part 'A' and 'B' and include both parts.}
\label{rot_14_16}
\end{figure}

To learn how strongly the magnetic field of the sunspot is twisted, we calculated the difference between 
the azimuth angle of the magnetic field in the photosphere and that of a potential field,
$\phi^{\prime}_{\rm{Ph}}$ -- $\psi_{\rm{Ph}}$,  where $\phi^{\prime}_{\rm{Ph}}$ and $\psi_{\rm{Ph}}$ 
denotes the azimuth angle in the photosphere and the azimuth angle of the potential field is derived from 
the vertical component of photospheric magnetic field, respectively. The potential field is
calculated with the Fourier method \citep{Alissandrakis_1981,Gary_1989}.
We also calculated the twist of the upper chromospheric magnetic vector, 
$\phi^{\prime}_{\rm{Ch}}$ -- $\psi_{\rm{Ch}}$, where $\phi^{\prime}_{\rm{Ch}}$
denotes the azimuth angle in the upper chromosphere and $\psi_{\rm{Ch}}$ represents the azimuth angle 
of the potential field calculated from the vertical component of the 
chromospheric magnetic field. 
Finally, we determined the differential twist between the photosphere and upper
chromopshere, i.e., the difference of the azimuth angle between the upper chromosphere
and photosphere, $\phi^{\prime}_{\rm{Ch}}$ -- $\phi^{\prime}_{\rm{Ph}}$. All twist maps 
are shown in Fig.~\ref{rot_14_16} for both days of observations. 
Positive values denote counterclockwise twist.
The sunspot obviously exhibits twist in both directions, but 
clockwise twist (green and blue in Fig.~\ref{rot_14_16})
are more common in both atmospheric layers on both days. Figs.~\ref{rot_14_16}(a),(b),(d), and (e) suggest a field 
of the sunspot diverging slightly away from the negative $y$-axes on 14 November 2010 and away from roughly 
the direction of the disk center on 16 November 2010.
The maps of $\phi^{\prime}_{\rm{Ch}}$ -- $\phi^{\prime}_{\rm{Ph}}$ show differential twists in both directions in some 
areas, but $\phi^{\prime}_{\rm{Ch}}$ -- $\phi^{\prime}_{\rm{Ph}}$ is small in most parts of the penumbra. 
Maps of $\phi^{\prime}_{\rm{Ch}}$ -- $\phi^{\prime}_{\rm{Ph}}$ also show that the differential twist is 
higher in the outer penumbra compared to inner penumbra.

\begin{figure}
\centering
      \includegraphics[width = 0.48\textwidth]{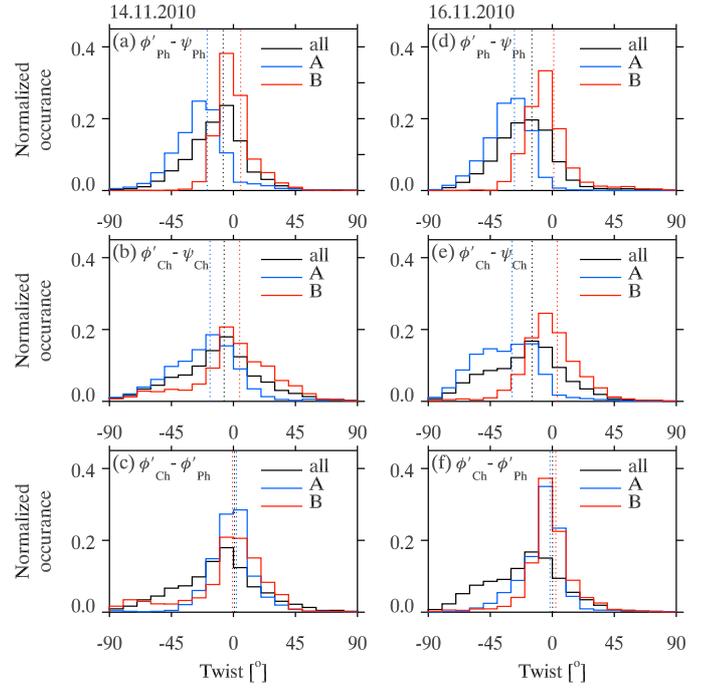}
      \caption{Histograms of the twist and differential twist of the magnetic field of the sunspot,
                with the differential twist denoting the difference between the azimuthal direction 
                of the field  in the photosphere, 
                $\phi^\prime_{\rm Ph}$ and chromosphere,  $\phi^\prime_{\rm Ch}$.  
                Panels (a) and (d) represent $\phi^{\prime}_{\rm{Ph}}-\psi_{\rm{Ph}}$.
                Panels (b) and (e) represent $\phi^{\prime}_{\rm{Ch}}-\psi_{\rm{Ch}}$.
                Panels (c) and (f) represent $\phi^{\prime}_{\rm{Ch}}-\phi^{\prime}_{\rm{Ph}}$.
                Blue, red, and black histograms are derived from pixels in the right, left,
                and all parts of the sunspot, respectively (the right and left
                part is separated by a line in Fig.~\ref{rot_14_16}). Dotted vertical lines
                represent the average twist. Panels in the left and right columns correspond to the observations recorded 
                on 14 November 2010 and 16 November 2010, respectively.}   
\label{hist}
\end{figure}

Fig.~\ref{rot_14_16} suggests that both senses of the twist are present in the observations of both days.
To better quantify this fact we divide the spots in two halves (identified as part-A and part-B) with the divisions line selected to
maximize the difference in the twist ($\phi^{\prime}_{\rm{Ph}}$ -- $\psi_{\rm{Ph}}$, solid black lines in Fig.~\ref{rot_14_16}).
Fig.~\ref{hist} shows histograms of the twists for the same panels as in Fig.~\ref{rot_14_16}, 
individually for both halves of the sunspot. On average part-A has clockwise twist and part-B has counterclockwise twist at both 
atmospheric heights on both days. 
On 14 November 2010, the average twist of the photospheric and 
upper chromospheric azimuth angle was  $-6^\circ$,
and  $-7^\circ$, respectively (see Table~\ref{tab6.4}). The average twist  
is  $-13^\circ$ in the photosphere and  $-12^\circ$
in the upper chromosphere on 16 November 2010. Histograms of twist maps indicate that the distribution 
of the upper chromospheric twist is wider than that of the photosphere, although a part of the excess 
may be due to the higher noise in the chromospheric values. On average, 
the differential twist between the photosphere and the chromosphere is very small
with values between $-1^\circ$ and $3^\circ$.

\begin{figure}
\centering
      \includegraphics[width = 0.38\textwidth]{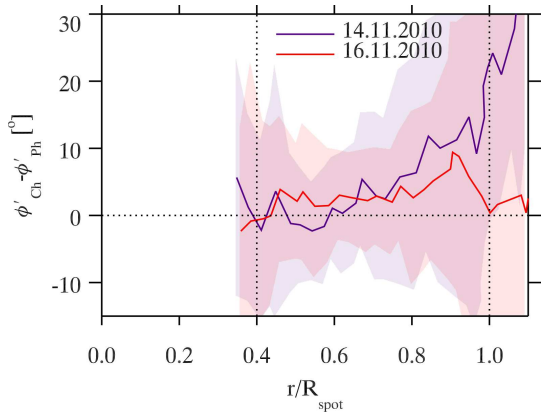}
      \caption{Radial dependence of $\phi^{\prime}_{\rm{Ch}}-\phi^{\prime}_{\rm{Ph}}$ as a function of 
               $r/R_{\rm{spot}}$. The blue and red curves correspond to $\phi^{\prime}_{\rm{Ch}}-\phi^{\prime}_{\rm{Ph}}$
               as obtained from the observations recorded on 14 November 2010 and 16 November 2010,
               respectively. Only the regular parts of the sunspots to the lower right of 
               the red lines in Figs.~\ref{radial_14}(a) and \ref{radial_16}(a) are considered.}     
\label{radial_rot}
\end{figure}

The radial dependence of the differential twist, $\phi^{\prime}_{\rm{Ch}}$ -- $\phi^{\prime}_{\rm{Ph}}$
for both days is presented in 
Fig.~\ref{radial_rot}. Azimuthal averages are only calculated for 
areas shown below the red lines in Fig.~\ref{radial_14}(a) and Fig.~\ref{radial_16}(a)
for 14 and 16 November 2010, respectively. On 14 November 2010  the differential twist increases from 
0\textdegree\, to $\sim$21\textdegree\, between the umbra-penumbra 
boundary and the outer boundary of the sunspot. On 16 November 2010 the differential twist also
increases outward through most of the penumbra, but then abruptly drops to near zero close to $R_{\rm{spot}}$. 
The average differential twist is on average $\sim$4\textdegree\, out to $r = R_{\rm{spot}}$. Hence, although
the average differential twist is always very small, this may be masked by the fact that the inner penumbra
displays a very small differential twist.

\section{Discussion and conclusions} \label{sec_4}
\subsection{Magnetic field strength and its vertical gradient} \label{sec_4.1}

We have measured and compared the magnetic field vector of a sunspot in its
photospheric and upper chromospheric layers. In the umbra we found the maximum value of
$B$ to be around 2.5/2.8\,kG\, in the photosphere and 1.6/1.8\,kG\, in the upper chromosphere from observations
recorded on 14 and 16 November 2010. On average the upper chromospheric umbral magnetic 
field strength is reduced by a factor 1.30-1.65 compared to the photosphere.    
These differences between the upper chromospheric and the photospheric magnetic 
field strength are comparable with the results of \citet{Rueedi_1995b} and \citet{Schad_2015}, 
who used the same spectral lines analyzed in this paper. A striking feature of the upper chromospheric magnetic field is that it is almost constant
from the center of the sunspot to the umbra-penumbra boundary; i.e., it decreases by less than a factor of 1.1.
In the photosphere it decreases from the center of the sunspot to the umbra-penumbra boundary
by a factor of 1.2-1.4, which is consistent with
values found earlier \citep{Solanki_1992b,Balthasar_1993,Skumanich_1994,Keppens_1996,
Westendrop_2001a,Mathew_2003,Borrero_2011, Tiwari_2015}.\footnote{We inferred
this number from published plots, or deduced it from numbers provided by the authors, for those papers where 
this number was not explicitly stated.}

The ratio of the magnetic field strength between the photosphere
and chromosphere drops from a factor of 1.4 in the umbra and the 
umbra-penumbra boundary steadily over the whole penumbra to reach a value of 1.0
at the outer boundary of the sunspot. We see a magnetic canopy structure 
outside the visible boundary of the sunspot, in the sense that the upper chromospheric field 
is higher by up to $\sim$300\,G compared to the field in the photosphere.
This magnetic canopy results from an expansion of the magnetic field of the sunspot with height
beyond its visible boundary as seen 
in continuum images. Results of the height-dependent inversions of the 
\ion{Si}{i}\, and \ionn{Ca}{i}\, lines, as discussed in \citetalias{Joshi_2017}, show that the base of the canopy 
lies in the photosphere. Sunspot magnetic canopies have been observed by \citet{Jones_1982} by
comparing magnetograms obtained in the photosphere using the \ionn{Fe}{i}\,8688\,\AA\, 
line with those obtained from the chromospheric \ionn{Ca}{ii}\,8542\,\AA\, line. 
Sunspot magnetic canopies in the photosphere have been regularly detected using different
spectral lines as a diagnostic \citep{Solanki_1992b, Lites_1993, Adams_1993, Solanki_1994, Skumanich_1994,
Keppens_1996,Rueedi_1998,Solanki_1999,Tiwari_2015}. 

We find that in the umbra, the magnetic field strength decreases 
from the photosphere toward the upper chromosphere at a rate of 
0.5-0.9\,G\,km$^{-1}$. These values are higher than the values of  0.4-0.6\,G\,km$^{-1}$ found by \citet{Rueedi_1995b}.
The reason for this discrepancy lies mainly in the assumption about the formation height 
of the \ionn{He}{i} triplet. Whereas \citet{Rueedi_1995b} assumed that in sunspots the
\ionn{He}{i} triplet forms 1500-2000\,km\, above the photosphere, 
we considered a height difference of 1000\,km following \citet{Centeno_2008}. 
\citet{Schad_2015} report a vertical gradient of the magnetic field strength around 0.5\,G\,km$^{-1}$
in the umbra. They also used the \ionn{Si}{i}\,10827.1\,\AA\, and \ionn{He}{i}\,10830\,\AA\, triplet
in their study and assumed a 1000\,km difference in formation height of these lines.  

In the penumbra the magnetic field strength decreases more slowly with 
vertical gradients between 0 and 0.7\,G\,km$^{-1}$. In the outer penumbra and outside the visible 
boundary of the sunspot the gradient has negative values, indicating the presence of a magnetic canopy.
Here the vertical gradient in the penumbra may be underestimated because of the underestimation
of the magnetic field strength in the photosphere due to unresolved opposite polarities \citep[see   ][]{Zakharov_2008,Scharmer_2013,Ruiz_2013,Franz_2013,
Tiwari_2013,Vannoort_2013,Joshi_2014,Joshi_2017}.

If we consider the difference in formation heights to be 1250\,km, as estimated by \citet{Joshi_2016}, then
the vertical gradient of the magnetic field in the umbra and penumbra would be 0.4-0.7\,G\,km$^{-1}$ and 0.0-0.6\,G\,km$^{-1}$,
respectively. 

The lightbridge exhibits $\sim$1\,kG\, lower magnetic field strength compared to the surrounding 
umbra at photospheric layers present on 14 November 2010. This is a well-known property of lightbridges
\citep{Rimmele_1997,Berger_2003,Jurcak_2006, Rimmele_2008,Sobotka_2013,Lagg_2014}. 
Signatures of the lightbridge are visible in the \ionn{He}{i}\,10830\,\AA\, spectrum, 
for example, the higher values of the line depression in the lightbridge compared
to those in the surrounding umbra (see Fig. 1(c)). However, the magnetic field in the upper 
chromosphere does not seem to be influenced by the lightbridge; at this height, the magnetic field 
above the lightbridge is not distinguishable from the field in the surrounding umbra. 
This result is consistent with the findings of \citet{Rueedi_1995b}, but is in contrast to 
that of \citet{Schad_2015}, who obtained a lower value of the magnetic field strength in the
chromosphere at the location of a lightbridge. Our results show that lightbridge is stronger by $\sim$200\,G\, 
compared to the photosphere. This and other results of our work indicate a 
cusp-like shape of the magnetic field lines above the lightbridge forming a canopy 
directly above the photosphere, which is completely covered by the umbral field at the chromospheric heights  
\citep[see also,][]{Jurcak_2006, Lagg_2014, Felipe_2016}. In the deep photosphere, 
magneto-convection can lower the field in a lightbridge considerably. In the upper atmosphere the strong field from 
the surrounding umbra expands to fill the volume above the lightbridge. 

\subsection{Inclination} \label{sec_4.3}

We found that the magnetic field in the penumbra is less inclined in the upper chromosphere
compared to the photosphere by 10-20\textdegree/5-10\textdegree\, on 14 and 16 November 2010. 
Our results are in agreement with \citet{Joshi_2016}, who found the upper chromospheric 
magnetic field in the penumbra to be more vertical by $\sim$12\textdegree\, compared to
that in the photosphere. More vertical magnetic fields in penumbrae at higher photospheric
layers compared to the deep photosphere have already been reported by \citet{Westendrop_2001a,
Sanchez_cuberes_2005,Borrero_2011}. \citet{Tiwari_2015} have also found that in azimuthal averages, 
the magnetic field becomes more vertical with height in the photosphere everywhere 
in the sunspot. They obtained the height-dependent magnetic field vector using spatially coupled inversions 
\citep{Vannoort_2012,Vannoort_2013} from observations recorded by the Spectropolarimeter
of the Solar Optical Telescope \citep[SOT/SP;][]{Tsuneta_2008}.
\citet{Joshi_2016} also observed a more vertical magnetic field in the higher layers of the photosphere than in its lower layers in a sunspot penumbra using high spatial resolution data 
obtained with the 1.5-meter GREGOR telescope \citep{Schmidt_2012} using the GRIS spectropolarimeter \citep{2012AN....333..872C}.

A monolithic flux tube expanding with height should have a more vertical field in 
the higher layers at a given spatial pixel; this is consistent with our observation.
In addition to this, the more inclined magnetic field in the lower photospheric layer may be further explained by the
penumbral fine structure. \citet{Borrero_2008}, \citet{Tiwari_2013}, and \citetalias{Joshi_2017}
have shown that the more vertical field of spines expands and covers the relatively 
horizontal magnetic field lines (intra-spines or filaments) in general accordance with the geometry proposed 
by \citet{Solanki_1993a}. In the upper chromosphere, \citet{Joshi_2016} therefore see that the peak-to-peak variation in 
the magnetic field inclination due to spine and inter-spine structure is reduced to $10^{\circ}-15^{\circ}$ compared to 
$20^{\circ}-25^{\circ}$ in the photosphere, again in qualitative agreement with this picture. In 
the azimuthal averages, the field appears to be more horizontal in the deeper photosphere, resulting from the smearing of 
horizontal and vertical field lines. This effect could also play a role in the umbra due to the presence of
umbral dots and lightbridges harboring a more horizontal magnetic field than the surrounding 
umbra in the deeper layers of the photosphere \citep[see, e.g.,][]{Riethmueller_2013,Lagg_2014}.

\subsection{Twist of the magnetic field of the sunspot} \label{sec_4.4}

At different places, the sunspot exhibits different senses of twist of the magnetic field vector in the azimuthal direction with respect to
a potential field for both\ the photosphere and 
chromosphere. However, the clockwise twist dominates and the average twist angle is found to be 
$\approx6^\circ$ on 14 November 2010 in the photosphere and the upper chromosphere. 
On 16 November 2010 the twist increases to $\approx12^\circ$ in both the photosphere and 
upper chromosphere. A clockwise twist corresponds to negative magnetic helicity \citep{Tiwari_2009}. The observed sunspot therefore 
follows the helicity hemispheric rule \citep{Hale_1925,Hale_1927,Richardson_1941,Hagino_2004, Nandy_2006,
Bernasconi_2005, Pevtsov_2008}. Sunspots with both senses of twist at the same time have been 
observed earlier \citep{Tiwari_2009,Socas_2005,Tiwari_2015}. We found that the sunspot has on average a negligible
twist in azimuth angle between the 
chromospheric and photospheric magnetic field when considering the whole sunspot. However, locally the twist maps show significant 
values of both signs in the sunspot. The radial dependence of the twist between the chromosphere and photospheric 
magnetic field indicates an increase from the inner penumbra to the outer penumbra. Similar twist gradients have been 
reported by \citet{Tiwari_2015} in the photospheric layers of a sunspot. \citet{Peter_1996} demonstrated that the decreasing 
radial magnetic field leads to increasing twist under influence of the Coriolis force on the 
radial flows (the inverse Evershed flow) in chromospheric penumbrae and superpenumbrae.\\

\noindent
The present study has revealed a number of basic properties of the magnetic field in the upper chromosphere of a sunspot. 
The differences between the results on the two days at which the sunspot was studied suggest that it would be worthwhile
to observe and analyze a number of sunspots at the two layers to determine the commonality of their behavior and to learn 
of the full range that the properties of the upper chromospheric magnetic field in sunspots can cover. Recently, the work of \citet{Joshi_2016} has 
given intriguing signs of small-scale structure (spine and interspine) in the upper chromosphere, i.e., in a layer in which magnetoconvection, 
the driver of similar structure seen in the photosphere, cannot be the cause of the structure. The
higher spatial resolution now afforded by the GRIS instrument \citep{2012AN....333..872C} on the GREGOR telescope makes it 
attractive to redo a similar analysis as presented in the current paper, but applied to data with a higher spatial resolution. 

\begin{acknowledgements}
The German Vacuum Tower Telescope is operated by the Kiepenheuer-Institut f\"{u}r
Sonnenphysik at the Spanish Observatorio del Teide of 
the Instituto de Astrof\'{\i}sica de Canarias (IAC).
This work was partly supported by the BK21 plus
program through the National Research Foundation (NRF) funded by the Ministry of Education of Korea.
\end{acknowledgements}

\bibliographystyle{aa} 
\bibliography{30875_am}
\end{document}